\documentclass[aps,prl,showpacs,twocolumn,superscriptaddress]{revtex4-1}

\usepackage{amssymb}
\usepackage{graphicx}
\usepackage{amsmath}
\usepackage{color}
\usepackage{hyperref}

\def\jqmc{\mbox{$\vec{J}$-QMC}}
\def\jzqmc{\mbox{$J_z$-QMC} }
\def\hfqmc{\mbox{HF-QMC} }
\def\s{\sigma}

\begin{document}

\title{Rotationally-Invariant Exchange Interaction: The Case of Paramagnetic Iron}

\author{V.~I.~Anisimov}
\affiliation{Institute of Metal Physics, Russian Academy of Sciences,
620990 Yekaterinburg, Russia}
\affiliation{Ural Federal University, 620990 Yekaterinburg, Russia}
\author{A.~S.~Belozerov}
\affiliation{Institute of Metal Physics, Russian Academy of Sciences,
620990 Yekaterinburg, Russia}
\affiliation{Ural Federal University, 620990 Yekaterinburg, Russia}
\author{A.~I.~Poteryaev}
\affiliation{Institute of Metal Physics, Russian Academy of Sciences,
620990 Yekaterinburg, Russia}
\author{I.~Leonov}
\affiliation{Theoretical Physics III, Center for Electronic Correlations and Magnetism,
Institute of Physics, University of Augsburg, D-86135 Augsburg, Germany}

\date{\today}

\begin{abstract}
We present a generalization of the spin-fluctuation theory of magnetism
which allows us to treat the full rotational invariance of the exchange
interaction.
The approach is formulated in terms of the local density approximation
plus dynamical mean-field theory (LDA+DMFT), providing a systematic
many-body treatment of the effect of spin-density fluctuations.
This technique is employed to study the electronic and magnetic properties
of paramagnetic $\alpha$~iron.
Our result for the Curie temperature is in good agreement with experiment,
while the calculations with the Ising-type exchange interaction yield almost
twice overestimated value.
\end{abstract}

\maketitle
The theoretical description of metallic magnets, especially those 
containing transition metals, is one of the central problems in condensed matter 
physics. 
Even today, in view of the great technological importance of such materials,
a detailed understanding of their electronic, magnetic, and structural properties at finite 
temperatures remains problematic.
This is mostly due to the presence of local magnetic moments above the magnetic ordering 
temperature which complicates the problem considerably and reduces the predictive 
power of first-principles calculations.
Various properties of metallic magnets can be understood by using the spin-fluctuation 
theory \cite{Moriya85} with its most general form based on a functional 
integral formulation \cite{Hubb_Strat}.
This formulation was employed to describe the formation
of local moments in paramagnetic metals~\cite{Wang69}    
by reducing the many-body problem 
to a one-particle problem in a fluctuating external magnetic field
and then evaluating the functional average.
Stimulated by these results, different analytical and numerical methods
have been developed, e.g. the well-known quantum Monte Carlo techniques \cite{Blankenbecler,Hirsch}.
By taking into account fluctuation corrections to the mean-field approximation, the spin-fluctuation 
theory have shown to provide a good qualitative description of the Curie-Weiss law behavior of magnetic 
susceptibility.
However, applications of this technique to describe, e.g., the $\alpha$-$\gamma$ phase transition in 
iron, do not lead to satisfactory results \cite{HP83}. 
In particular, it predicts the bcc-fcc phase transition to occur below the Curie temperature, $T_C$, while, 
in fact, this phase transition occurs 150~K above $T_C$.

%
%
%
%
%

The LDA+DMFT method \cite{LDA+DMFT}, a combination of the \textit{ab~initio} 
local density approximation (LDA) of the density functional theory and dynamical
mean-field theory (DMFT), nowadays has become a state-of-the-art
approach for the calculation of the electronic and magnetic properties 
of correlated electron compounds \cite{Anisimov_book10}.
%
%
Applications of the LDA+DMFT to study transition metal compounds have 
shown to give a good quantitative description of localized as well as 
itinerant electron states \cite{Licht01,Kunes_Belozerov,Katanin10}.
These calculations predict the correct values
of the local magnetic moment and magnetization,
while the magnetic transition temperature turns
out to be significantly overestimated.
It was proposed that this is caused by the 
single-site nature
of the DMFT approach, which is not able to capture the reduction
of magnetic transition temperature due to long wavelength spin waves \cite{Licht01}.
Nevertheless, the substantial overestimation of the magnetic transition temperature
was also observed in other compounds, even in those where non-local
spin fluctuations are negligible~\cite{Kunes_Belozerov}.
This implies that the approximate form of the local Coulomb repulsion,
restricted to the Ising-type exchange interaction,
can be the reason of this systematic overestimation.
The recently proposed continuous-time quantum Monte Carlo algorithms~\cite{CT-QMC}
as well as some other quantum impurity solvers~\cite{other_solvers}
allow one to treat the Coulomb interaction in its general form
retaining rotational symmetry of spin.
However, applications of these techniques so far have been limited to
simple 
model systems due to the high computational costs.
Therefore, only the density-density part of the Coulomb repulsion,
that implies the Ising form of the exchange interaction,
has been employed in the most material specific calculations
of correlated electron materials.
However, as we will show below on the example of paramagnetic iron,
the 
retaining of spin rotational symmetry is crucial
for the correct description of the magnetic properties.

In this Letter, we present the spin-fluctuation theory of magnetism
which is formulated in the framework of the LDA+DMFT method. 
The approach provides a systematic treatment of the effect
of local electronic correlations by reducing the many-body problem
to the functional integral over a fluctuating magnetic field
on an effective impurity. 
The spin-fluctuation theory is generalized by replacing a scalar
fluctuating magnetic field to a vector one.
This allows one to take into account the full rotational invariance
of the exchange interaction instead of the approximate Ising-type form.
The proposed method is employed to study the electronic and magnetic 
properties of paramagnetic $\alpha$~iron, resulting in
the Curie temperature value which is in good agreement with experiment.


We start with the simple Hamiltonian of the Coulomb interaction in
the following form
\begin{eqnarray} \label{UJ}
\widehat{H}_{\textrm{Coul}}
=\frac{1}{2}\sum_{\mu,\nu,\s}U\widehat{n}_{\mu\s}\widehat{n}_{\nu\bar{\s}}
+\frac{1}{2}\sum_{\scriptstyle \mu,\nu,\s \atop \scriptstyle \mu\neq\nu}
(U-J)\widehat{n}_{\mu\s}\widehat{n}_{\nu\s}, \quad
\end{eqnarray}
where $\widehat{n}_{\mu\s}$ denotes the electron number operator
with the spin~$\s$~(${=\uparrow,\downarrow}$) at the orbital~$\mu$.
Using the total electron number operator,
$\widehat{N}=\sum_{\mu\s}\widehat{n}_{\mu\s}$,
and the $z$-projection of the spin operator,\,
${\widehat{S}_z
=\sum_{\mu} (\widehat{n}_{\mu\uparrow}-\widehat{n}_{\mu\downarrow})}/2$,\, 
the Hamiltonian can be rewritten as
\begin{eqnarray} \label{UJ2}
\widehat{H}_{\textrm{Coul}} = \frac{1}{2}\bar{U}\widehat{N}
(\widehat{N}-1)+\frac{1}{4}J\widehat{N}-J\widehat{S}_z^2,
\end{eqnarray}
where $\bar{U}=U-J/2$\, is the average value of the Coulomb interaction.
This Hamiltonian represents the density-density part of the Coulomb
interaction and contains the exchange interaction in the Ising-type form.
To restore the spin rotational symmetry, one should
replace the $z$-projection of the spin operator, $\widehat{S}_z$, to
the vector spin operator, $\widehat{\vec{S}}$.
Therefore, the Hamiltonian with the rotationally-invariant exchange interaction reads
\begin{eqnarray} \label{UJ2I2}
\widehat{H}_{\textrm{Coul}}=\frac{1}{2}\bar{U}\widehat{N}(\widehat{N}-1)
+\frac{1}{4}J\widehat{N}-J\widehat{\vec{S}}^2.
\end{eqnarray}
Following the spin-fluctuation theory, 
we neglect the charge fluctuations but preserve
the spin dynamics (magnetic moment fluctuations).
Employing the static mean-field approximation for the first 
term in Eq.~(\ref{UJ2I2}) and introducing the double counting correction for 
the Coulomb interaction, the Hamiltonian of the system can be expressed as
\begin{eqnarray} \label{UJ4}
\widehat{H} = \widehat{H}_\textrm{LDA}+\bar{U}(n_d-n_{d0})\widehat{N}
-J\widehat{\vec{S}}^2,
\end{eqnarray}
where $\widehat{H}_\textrm{LDA}$ is the LDA Hamiltonian,
${n_d=\langle\,\widehat{N}\,\rangle}$ is the average number of $3d$ electrons,
and $n_{d0}$ is the LDA value for ${n_d}$.

In the DMFT approach the lattice problem with the Hamiltonian (\ref{UJ4})
is mapped onto a quantum impurity model. 
Using the general form of the Hubbard-Stratonovich transformation~\cite{Hubb_Strat},
the partition function can be expressed as a functional integral
\begin{eqnarray} \label{JQ3I} 
Z =\int \textsl{D}\vec{\xi}(\tau) \, \textrm{exp}[-\frac{\pi}{\beta}
\int^{\beta}_{0} \vec{\xi}^{\,2}(\tau) d\tau] Z(\vec{\xi}),
\end{eqnarray}
where
\begin{eqnarray} \label{part_funct} 
Z(\vec{\xi}) = & \textrm{Tr}\{\,\textrm{T}_{\tau} \, \textrm{exp}
[-\beta\widehat{H}_\textrm{LDA}-\beta \bar{U}(n_d-n_{d0})\widehat{N} \nonumber \\
& + 2c \int^{\beta}_{0} \vec{\xi}(\tau)\widehat{\vec{S}} d\tau]\}.
\end{eqnarray} 
Here, $\textrm{T}_\tau$ denotes the time ordering operator,
$\beta$ the inverse temperature, and ${c=\sqrt{\pi J/\beta}}$.
Function $\vec{\xi}(\tau)$ stands for an effective magnetic field
resulting in the potential
$\widehat{V}(\tau)= 2 c \, \vec{\xi}(\tau) \widehat{\vec{S}}$.
The functional integral over all fluctuating fields
gives a solution of the impurity problem.

In the functional integral formulation of the conventional spin-fluctuation theory,
the fluctuating magnetic field in Eq.~(\ref{JQ3I}) is considered to be scalar.
The generalization to a vector field corresponds to the transition from
Eq.~(\ref{UJ2}) to Eq.~(\ref{UJ2I2}) and
allows one to take into account the spin-rotational symmetry,
thereby extending the theory from the Ising-type exchange interaction
to the full rotationally-invariant one.
%
%
Dividing the imaginary time interval $[0,\beta]$ on $L$ slices
of length $\Delta\tau$ and using the Trotter breakup
for the exponential operator in Eq.~(\ref{part_funct}),
the partition function~$Z(\vec{\xi})$ for a given~$\vec{\xi}(\tau)$
can be expressed as
\begin{eqnarray} \label{trotter} 
Z(\vec{\xi}) \simeq  \textrm{Tr} \left\{ \textrm{T}_{\tau} \prod_{l=1}^L
( \textrm{exp}[-\Delta\tau\widehat{H}_0] \:
\textrm{exp} [ \widehat{V}(\tau_l) ] ) \right\},
\end{eqnarray} 
where $\widehat{H}_0 = \widehat{H}_\textrm{LDA} + U(n_d-n_{d0})\widehat{N}$
is the $\vec{\xi}$-independent part of the Hamiltonian.
These equations are similar to those of the 
Hirsch-Fye quantum Monte Carlo (HF-QMC) method \cite{Hirsch}.
The partition function can be written as
%
\begin{eqnarray} 
\label{JP2I} 
Z& =& \sum_{\{\vec{\xi}\}}
\textrm{exp}[-\frac{\pi}{L}\sum_{l=1}^L \vec{\xi}^{\,2}(\tau_l)]
\prod_{\mu} \textrm{det}[G_{\mu}^{-1}(\vec{\xi})].
\end{eqnarray} 
However, instead of a single spin-flip as in the HF-QMC method,
here one should stochastically change the value of the 
field $\vec{\xi}(\tau)$ for a random value of imaginary time.
Due to the rotational symmetry of the exchange interaction,
the interacting Green function becomes non-diagonal in spin indexes.
%
%
%
The computational scheme where the partition function is calculated
with an auxiliary vector magnetic field is referred below as $\vec{J}$-QMC.
By taking into account only the $z$-component of the field, the approximate form
of the local Coulomb interaction, limited to the Ising-type exchange interaction,
is assumed (referred as $J_z$-QMC).
%


Elemental iron is one of the most famous itinerant-electron ferromagnets
which exhibits localized moment behavior above the Curie temperature, $T_C$.
Although various properties of the low-temperature ferromagnetic state 
of Fe can be understood within the density-functional theory \cite{Fe_LDA},
applications of these techniques to describe the {\it paramagnetic} state do not lead to 
satisfactory results.
Clearly, an overall understanding of the properties of iron requires a 
formalism which takes into account the existence of local magnetic moments above $T_C$ \cite{Hubbard_Fe}.
%
%
Recent applications of the LDA+DMFT have shown to provide
a qualitatively correct description of the electronic,
magnetic, and structural properties of paramagnetic iron
\cite{Katsnelson99,Katsnelson00,Licht01,Katanin10,Fe_Leonov}. 
However, a quantitative agreement has been achieved
only in terms of the reduced temperature $T/T_C$,
while the calculated Curie temperature~\cite{Licht01} was found to be about
twice larger than the experimental value of 1043~K \cite{susc_exp}.
%
%
The average Coulomb interaction in the Fe $3d$ shell is considerably smaller
than the bandwidth, showing no evidence for the formation of Hubbard bands
in the spectral function.
However, due to the strong exchange interaction~\cite{Katanin10},
the local magnetic moments are formed that is accompanied
by the loss of coherence for the metallic states due to the scattering
of electrons on the fluctuating spins.
The charge fluctuations in paramagnetic iron are of high frequency
with the $3d$ electrons being far from the localization limit.
These arguments make iron an ideal candidate for our study.
%
%
%
%
%
%
%

To calculate the electronic structure of paramagnetic $\alpha$~iron
within the LDA, the tight-binding linear muffin-tin orbital
(TB-LMTO) method was employed \cite{LMTO}.
The low-energy Hamiltonian containing the $4s$, $4p$, and $3d$ states
has been constructed with use of the
$N$th-order muffin-tin orbital ($N$MTO) method~\cite{Andersen00}.
In our calculations, we used the value of the screened Coulomb interaction,
${U = 2.3}$~eV, and the value of Hund's exchange, ${J = 0.9}$~eV,
which are consistent with the previous estimations~\cite{Katsnelson99,Licht01,Coco05,Katanin10}.

\begin{figure}
\centering
\includegraphics[clip=true, width=0.45\textwidth]{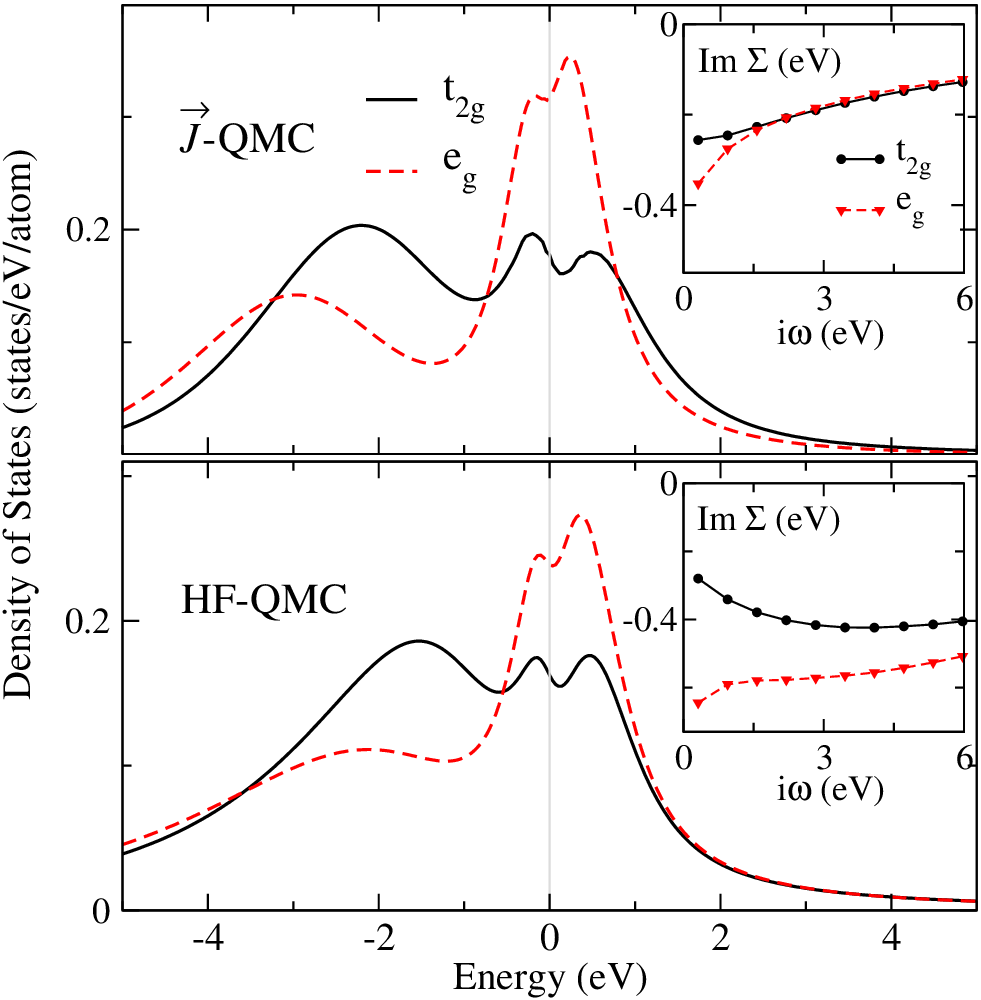}
\caption{(Color online) Partial densities of states obtained
         by the \jqmc\ (top panel) and \hfqmc (bottom panel) calculations within LDA+DMFT.
         The Fermi level is indicated by the vertical (gray)
         line at zero energy.
         Insets: imaginary parts of the self-energies.
         \label{fig:dos}}
\end{figure}

In Fig.~\ref{fig:dos} we present the partial densities of states
and the corresponding imaginary parts of the self-energies obtained by the LDA+DMFT
at $\beta = 10$~eV$^{-1}$.
The \hfqmc and \jqmc\ calculations give quantitatively similar results,
reproducing the splitting in the density of states of the $e_g$-orbitals
near the Fermi level caused by exchange interaction~\cite{Katanin10}.
The splitting in the density of states of the $t_{2g}$-orbitals
is found to occur in the LDA calculation and hence can be attributed 
to the band-structure effects.
In both approaches, the self-energies for the $t_{2g}$ orbitals
remain Fermi-liquid-like,
while the ones for the $e_g$ orbitals diverge at low frequencies.
%
The latter indicates the formation of local magnetic moments that implies
more incoherent $e_g$ states and itinerant $t_{2g}$ states~\cite{Katanin10}.
%
%
We note that at low frequencies the self-energies obtained
by the \jqmc\ method are close to those of the HF-QMC.
This indicates that the physics near the Fermi level is dominated
by the spin fluctuations while the charge fluctuations play a minor role.
%

\begin{figure}
  \centering
  \includegraphics[clip=true, width=0.45\textwidth]{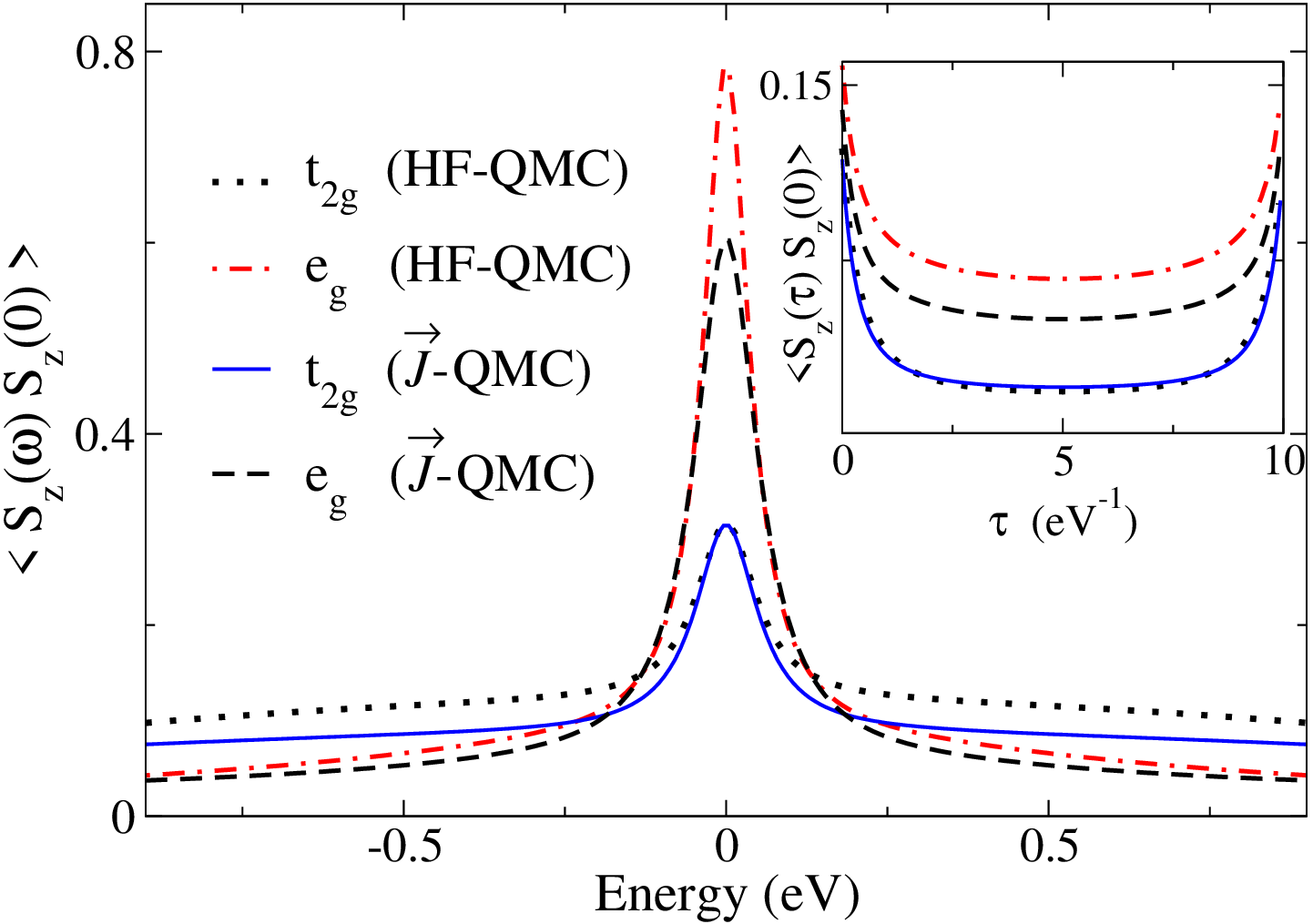}
  \caption{(Color online) Spin-spin correlation functions 
            on the real and imaginary energy (inset) axes calculated
            by the \hfqmc and \jqmc\ within LDA+DMFT. 
           \label{fig:sisj}}
\end{figure}

In Fig.~\ref{fig:sisj} we show our results for the orbitally-resolved spin-spin 
correlation functions on the real and imaginary energy axes.
In both approaches, the non-Fermi-liquid behavior of the $e_g$ electrons
yields a pronounced peak at zero energy of the real energy axis indicating
the presence of local magnetic moments.
%
%
The satisfactory agreement of the results obtained by the \hfqmc and \jqmc\ methods 
suggests that the effect of the charge fluctuations, neglected within the \jqmc\ approach, 
is minor.

\begin{figure}[t]
\centering
\includegraphics[clip=true, width=0.45\textwidth]{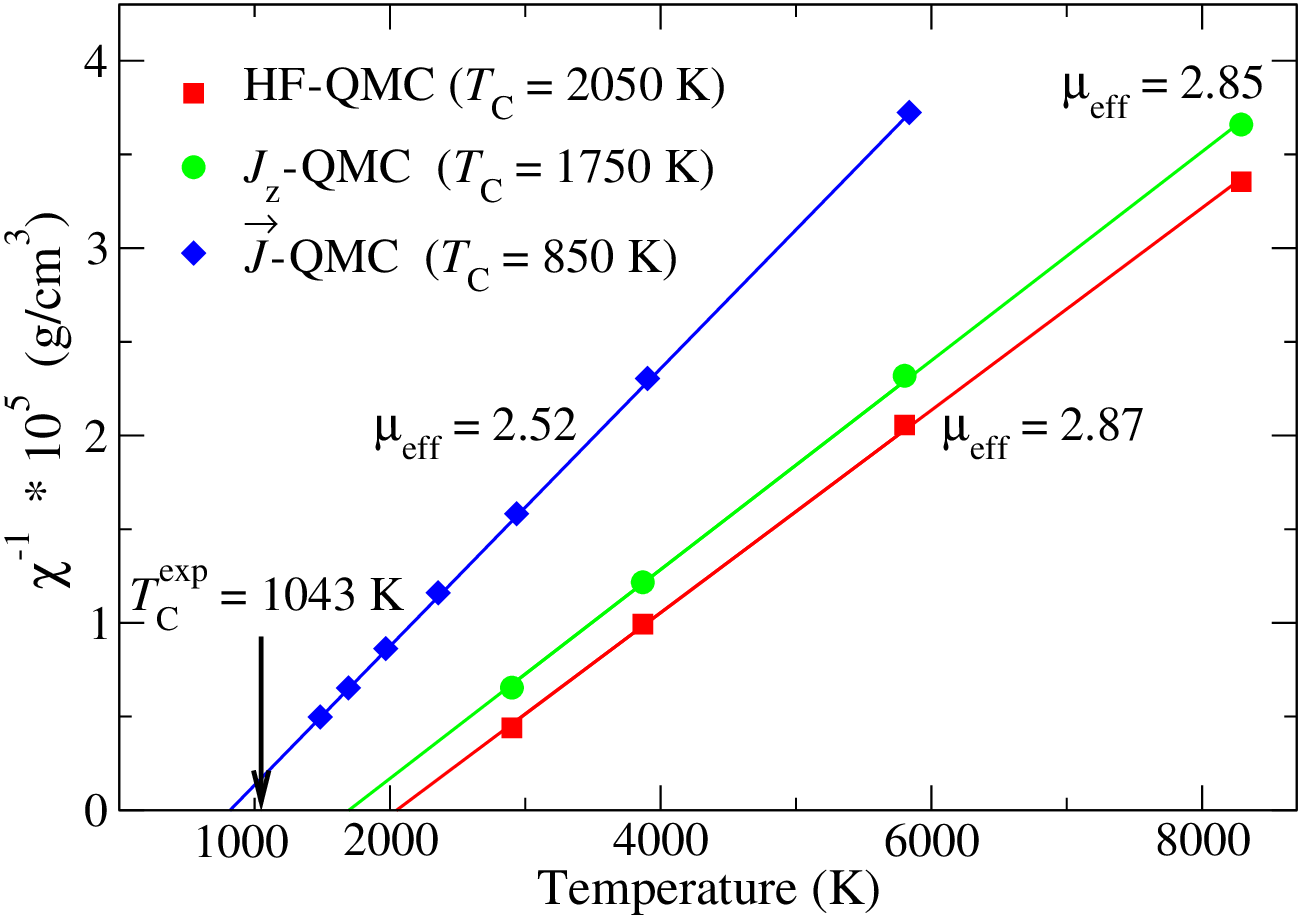}
\caption{(Color online) Temperature dependence of the inverse
         uniform magnetic susceptibility obtained by the LDA+DMFT.
         The straight lines depict the least-squares fit to the Curie-Weiss law.
         The experimental value of $T_C=1043$~K
         is denoted by (black) arrow.
         The experimental value of the local magnetic moment 
         is $\mu_\textrm{eff}^\textrm{exp} = 3.13~\mu_B$ \cite{susc_exp}.
         \label{fig:susc}}
\end{figure}

To proceed further we compute the uniform magnetic susceptibility
as a response to an external magnetic field.
The temperature dependence of the inverse uniform magnetic susceptibility
obtained by the LDA+DMFT shows a linear behavior at high temperatures
(Fig.~\ref{fig:susc}).
This indicates the presence of local magnetic moments and corresponds
to the Curie-Weiss law, ${\chi^{-1}=3(T-T_C)/\mu_{eff}^2}$, where
$T_C$ is the Curie temperature, $\mu_{eff}$ the effective local magnetic moment.
The results of the least-square fit to the Curie-Weiss law 
are shown in Fig.~\ref{fig:susc} by the straight lines.
%
%
%
%
%
It is clearly seen that the \hfqmc method, limited to the Ising-type
exchange interaction, overestimates the Curie temperature value almost twice.
The \jzqmc approach, which has the Ising-type exchange interaction,
gives a slightly smaller value of the $T_C$ than the \mbox{HF-QMC}.
This confirms the validity
of the static approximation for the charge degrees of freedom.
Taking into account the full rotationally-invariant exchange interaction,
our calculations result in a substantial decrease of the $T_C$ value,
which is now found to be in satisfactory agreement with experiment.
%
%
These findings are compatible with the results of 
the recent two-band model studies \cite{2band_models}.


In conclusion, we presented a generalization of the spin fluctuation-theory
of magnetism which allows one to take into account the full rotational invariance
of the exchange interaction.
The approach is formulated in terms of the LDA+DMFT method,
providing a systematic many-body treatment of the effect
of spin-density fluctuations.
We employed this new technique to study the electronic and magnetic properties
of $\alpha$~iron.
%
%
%
Our results agree well with experiment and show that the overestimation
of the Curie temperature by LDA+DMFT is mostly related to the approximate (Ising-type) treatment
of the exchange Coulomb interaction rather than to the single-site nature of the DMFT.


\begin{acknowledgments}
The authors thank D.~Vollhardt, A.~Lichtenstein, A.~Rubtsov,
and A.~Millis for useful discussions.
%
%
This work was supported by the Russian Foundation for Basic Research 
(Projects Nos. 10-02-00046a, 12-02-91371-CT$\_$a),
the fund of the President of the Russian Federation for
the support of scientific schools NSH-6172.2012.2,
the Program of the Russian Academy of 
Science Presidium ``Quantum microphysics of condensed matter''.
Support by the Deutsche Forschergemeinschaft through TRR~80 and FOR~1346
is gratefully acknowledged.
\end{acknowledgments}

\end{document}